\documentclass[aps,prb,twoside,twocolumn,superscriptaddress,longbibliography]{revtex4-1}
\usepackage[utf8]{inputenc}
\usepackage{amsmath}
\usepackage{amssymb}
\usepackage{graphicx}
\usepackage{lipsum}
\usepackage{esint}
\usepackage{float}
\usepackage{times}
\usepackage[colorlinks,linkcolor=blue,anchorcolor=blue,citecolor=blue]{hyperref} 

\makeatletter

\usepackage{dcolumn}
\usepackage{bm}
\usepackage{xcolor}
\usepackage{babel}
\makeatother
\usepackage{babel}
\graphicspath{}

\begin{document}

\title{Chemical Rules for Stacked Kagome and Honeycomb Topological Semimetals}

\author{Liqin Zhou}\email{These authors contribute equally to this work.}
\affiliation{Beijing National Laboratory for Condensed Matter Physics and Institute of Physics, Chinese Academy of Sciences, Beijing 100190, China}

\author{Fazhi Yang}\email{These authors contribute equally to this work.}
\affiliation{Beijing National Laboratory for Condensed Matter Physics and Institute of Physics, Chinese Academy of Sciences, Beijing 100190, China}

\author{Shuai Zhang}
\affiliation{Institute of Theoretical Physics, Chinese Academy of Sciences, Beijing 100190, China}

\author{Tiantian Zhang}
\email{ttzhang@itp.ac.cn} 
\affiliation{Institute of Theoretical Physics, Chinese Academy of Sciences, Beijing 100190, China}


\begin{abstract}

We study the chemical rules for predicting and understanding topological states in stacked kagome and honeycomb lattices in both analytical and numerical ways. 
Starting with a minimal five-band tight-binding model, we sort out all the topological states into five groups, which are determined by the interlayer and intralayer hopping parameters. 
Combined with the model, we design an algorithm to obtain a series of experimentally synthesized topological semimetals with kagome and honeycomb layers, i.e., IAMX family (IA = Alkali metal element, M = Rare earth metal element, X = Carbon group element), in the inorganic crystal structure database. 
A follow-up high-throughput calculation shows that IAMX family materials are all nodal-line semimetals and they will be Weyl semimetals after taking spin-orbit coupling into consideration. 
To have further insights into the topology of the IAMX family, a detailed chemical rule analysis is carried out on the high-throughput calculations, including the lattice constants of the structure, intralayer and interlayer couplings, bond strengths, electronegativity, and so on, which are consistent with our tight-binding model. 
Our study provides a way to discover and modulate topological properties in stacked kagome and honeycomb crystals and offers candidates for studying topology-related properties like topological superconductors and axion insulators.

\end{abstract}
\maketitle

\section{Introduction}
\begin{figure}
  \centering
  \includegraphics[width=0.45\textwidth]{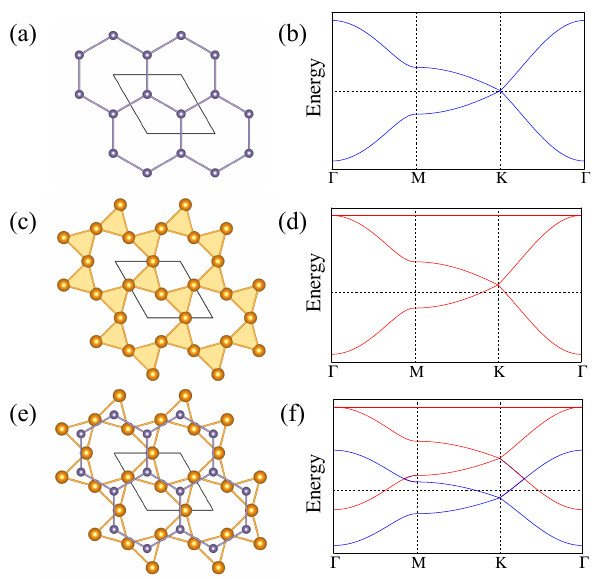}
  \caption{Schematic diagrams of three lattices and their band structures. (a-b) are for the honeycomb lattice, (c-d) are for the distorted kagome lattice, and (e-f) are for the system with stacked kagome and honeycomb lattice.\label{fig1}}
\end{figure}

The simplest model with topological phases arises from the honeycomb lattice, with the most famous material being graphene, 
in which the band structure features a Dirac cone at K and a van Hove singularity (vHs) at M~\cite{RevModPhys.81.109, Geim2007, Polini2013,PhysRevLett.106.116803,PhysRevLett.106.156402,PhysRevB.104.045139,PhysRevB.92.045108}, as shown in Figs.~\ref{fig1} (a-b).
Recently, the kagome lattice, a two-dimensional (2D) grid structure composed of corner-sharing triangles, has garnered significant attention in condensed matter physics due to its intriguing and diverse array of exotic phenomena~\cite{6.3.306,Jovanovic2022,Yin2022,Neupert2022,ptok2023phononic,PhysRevB.104.054305}, including charge density waves~\cite{PhysRevB.87.115135,PhysRevB.85.144402,PhysRevLett.127.046401,PhysRevLett.110.126405}, topological states~\cite{PhysRevB.80.113102,science.aav2873,science.aav2334,Yin2018,Yin2019,Ye2018}, and quantum spin liquids~\cite{yan2011spin,PhysRevX.6.041007,PhysRevB.77.224413,qute.202000126}. The kagome lattice shares similar properties in the band structure with the honeycomb lattice, including the Dirac cone and the vHs known as saddle point. Nonetheless, it also manifests distinct characteristics of flat band~\cite{PhysRevB.86.121105}, as depicted in Figs.~\ref{fig1} (c-d). 
Recently discovered kagome superconductors, such as CsV$_{3}$Sb$_5$, have also sparked curiosity regarding the potential of unconventional superconductivity and intertwined phase in kagome lattice~\cite{PhysRevLett.125.247002,Jiang2021,PhysRevX.11.031050, chen2021roton,Zhao2021,Hu2022,Li2022,mielke2022time,zheng2022emergent,zhong2023nodeless}. 

In the past, topological materials with kagome lattice and honeycomb lattice were studied separately, including nodal-line semimetals and Weyl semimetals~\cite{ozawa2019two,nie2017topological,zhang2017topological,zhang2022endless,lu2017two}, among which studies on kagome lattices were mainly focused on magnetic materials, such as magnetic Weyl semimetals Co$_3$Sn$_2$S$_2$~\cite{wang2018large,science.aav2873,science.aav2334} and Fe$_3$Sn$_2$~\cite{ye2019haas,yao2018switchable,PhysRevLett.125.076403}. Thus, the interplay between the kagome lattice and honeycomb lattice is still in the early stage and needs to be explored on both the analytical model level and materials level.

In this work, we first construct a minimal five-band tight-binding (TB) model to study the topology of stacked honeycomb and distorted kagome lattice. 
Then, based on the results of the TB model, we sweep through the inorganic crystal structure database (ICSD)~\cite{icsd2004} and discover 298 experimentally synthesized new ideal topological materials, i.e., IAMX family with the stacked distorted kagome and honeycomb lattice (IA = Alkali metal element, M = Rare earth metal element, X = Carbon group element)~\cite{zhang2019catalogue,PhysRevX.8.031069,song2018quantitative,vergniory2019complete,po2017symmetry}. The crystal structure of IAMX family is characterized by the hexagonal lattice with Fe$_2$P-type, with the space group of $P$-62$m$ (No. 189)~\cite{CzybulkaSteinbergSchuster,zaac19905800106}, and the interlayer interactions of the IAMX family are chemical bonds rather than the van der Waals forces comparing to the previous studied kagome materials. 
We perform a high-throughput calculation for IAMX compounds and discover four distinct topological node-ring configurations in the absence of spin-orbital coupling (SOC), which is closely associated with different chemical conditions, including the lattice constants of the structure, interlayer and intralayer interactions, bond strengths, electronegativity, and so on. Our numerical results match our TB model very well. 
Further calculation shows that LiNdGe is not a superconductor with $T_c<10^{-4}$ K under the BCS scenario, which implies IAMX family materials may be candidates for unconventional (topological) superconductors when the superconductivity is achieved by experiments. Our theory and findings offer a method for the stable realization of topological states, particularly Weyl semimetals, in the stacked kagome and honeycomb lattices. Combining the topological states with the charge instability arising from vHs in kagome material~\cite{PhysRevB.87.115135,PhysRevB.85.144402,PhysRevLett.127.046401,PhysRevLett.110.126405}, we provide an effective pathway for achieving charge density wave (CDW) in the Weyl semimetal state~\cite{PhysRevB.87.161107,PhysRevB.92.125141,PhysRevB.102.115159,gooth2019axionic,shi2021charge}. Our findings present an efficient approach for exploring the axion insulator state in kagome materials.

\begin{figure}[]
  \centering
  \includegraphics[width=0.45\textwidth]{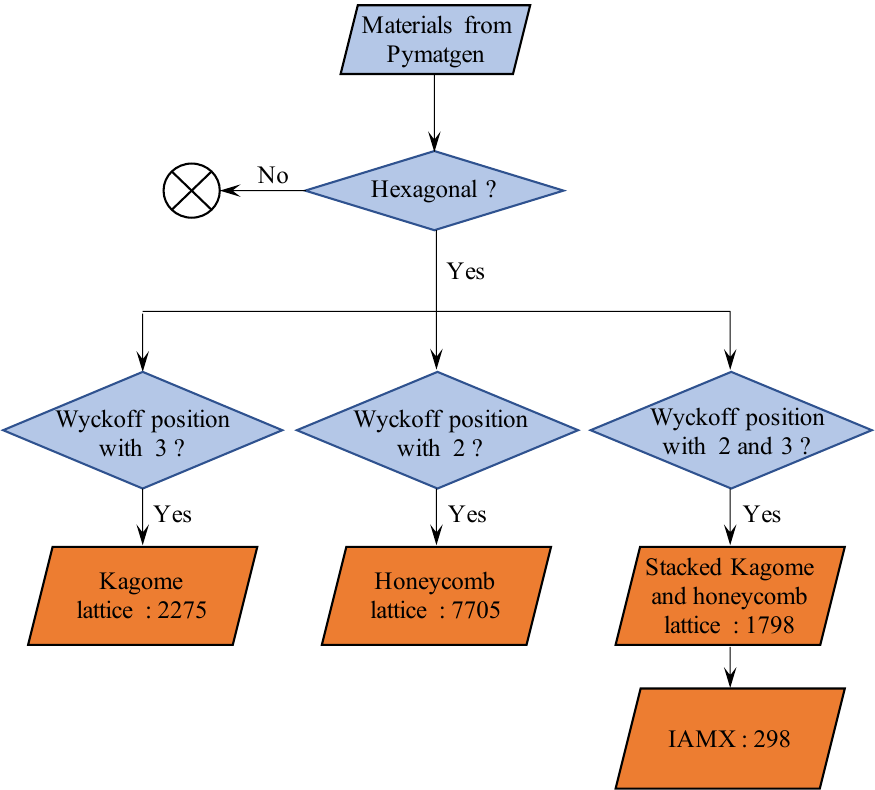}
  \caption{The flowchart for screening compounds with kagome lattice, honeycomb lattice, and their stacking lattice. First, the lattice should be a hexagonal one. Then, the crystal is classified based on their Wyckoff positions. If the Wyckoff position has 2? (? stands for different letters to distinguish different Wyckoff positions with the same site symmetry), the compound has an isolated honeycomb lattice; If the Wyckoff position has 3?, the compound has an isolated kagome lattice; If the Wyckoff positions have both 2? and 3?, the compound has the stacked structure of kagome and honeycomb lattice. We obtain 298 IAMX compounds with stacked lattices and ideal topological properties, which will be further proceeded with high-throughput calculations based on DFT.
\label{fig2}}
\end{figure}

\section{method}
\begin{figure*}
  \centering
  \includegraphics[width=1.0\textwidth]{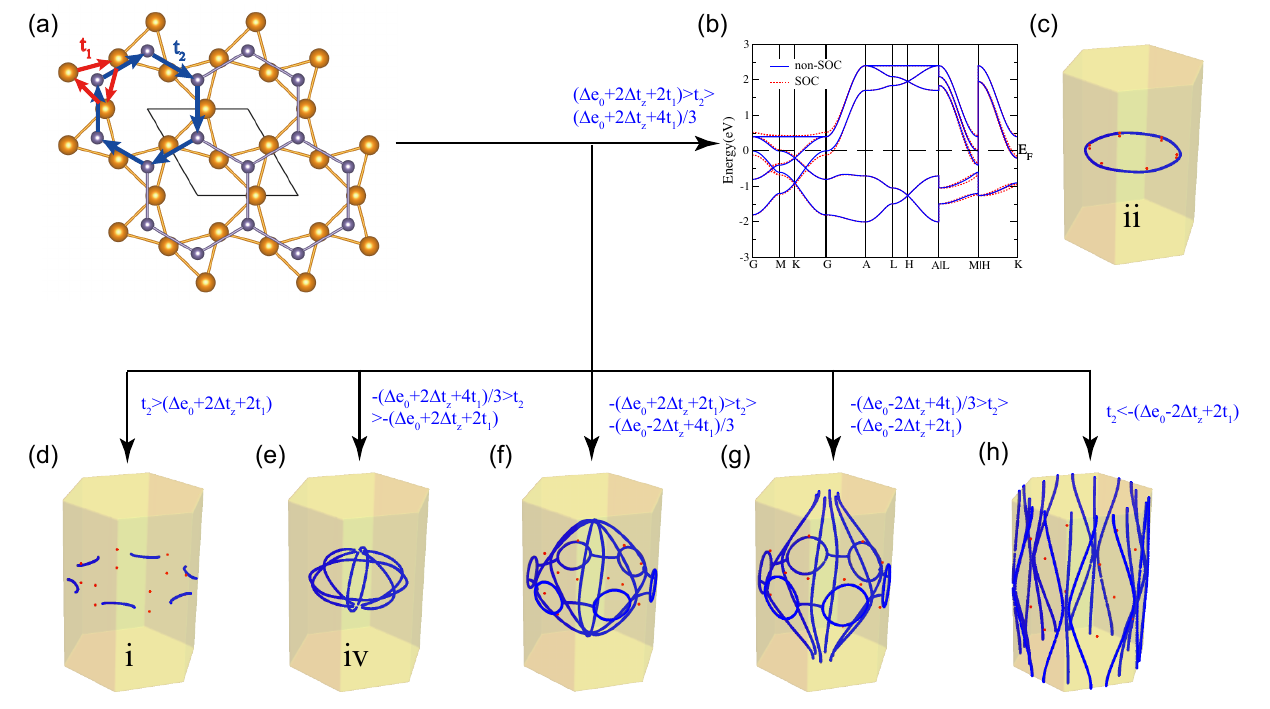}
  \caption{(a) Lattice model, (b) band structure, and (c-h) topological configuration for the stacked lattice tight-binding model. The blue lines represent the nodal ring in the non-SOC case, while the red dots represent the Weyl points in the SOC case. The shape of the nodal ring changes with various $t_2$, with $\Delta e_0 = e_1 - e_2, \Delta t_z = t_{z1} - t_{z2}$ as depicted in the figure. The results of tight-binding model with parameters $e_1 = 1, e_2 = -1, t_1 = -0.2, t_z = 0.2, t_{z1} = -0.5, t_{z2} = 0.05$, and $t_2 =$ 0.7 (d), 0.3 (b)-(c), -0.3 (e), -0.7 (f), -0.9 (g), -4 (h).
\label{fig3}}
\end{figure*}

The minimal five-band TB model for stacked kagome and honeycomb lattice without SOC is shown in Eq.~(\ref{eq1}), with $p_z$ orbital at Wyckoff position 1a, forming a honeycomb lattice, and $d_{z^2}$ orbital at Wyckoff position 3f, forming distorted kagome lattice. 
The model belongs to space group $P$-62$m$ (No.189) and with the next-next-neighbor hopping, as shown in Fig. ~\ref{fig3} (a)~\cite{zhang2022magnetictb}. 
Fig.~\ref{fig3} (c-h) are the node-ring configurations obtained in the tight-binding model without SOC with different parameters, as marked by the blue lines. After considering SOC, these nodal rings will evolve into Weyl points marked by red dots. 

The material screening process follows the flowchart shown in Fig.~\ref{fig2}. 
We first pick out all compounds with hexagonal lattices, then diagnose whether the compounds are stacked kagome and honeycomb lattices by the atomic Wyckoff positions. If atoms are located at Wyckoff positions of 3? or 2?, there will be a kagome and honeycomb sublattice. 
Here, 3? means there will be 3 symmetry-related atoms in the unit cell, and ? represents different letters for different Wyckoff positions with the same site symmetry. 
In this work, we discover 400 IAMX family candidates for DFT calculation, of which 298 are ideal topological materials for further study.

High-throughput DFT calculations for the IAMX family are performed by Vienna {\it ab initio} simulation package (VASP)~\cite{PhysRevB.48.13115}. The exchange-correlation functional is described by the generalized gradient approximation (GGA) with the Perdew-Burke-Ernzerhof (PBE) functional type~\cite{blochl1994projector,perdew1996generalized}. The cutoff energy for plane-wave basis is set to 520 eV and the BZ is sampled by 7$\times$7$\times$11 $\Gamma$-centered \emph{k} mesh. The structural optimization is performed until forces on each atom are less than 0.001 eV/$\AA$. To further obtain the topological information, we constructed the tight-binding model with \emph{d} orbitals of M atoms and \emph{p} orbitals of X atoms based on the maximally-localized Wannier functions (MLWF) method~\cite{mostofi2014updated}. The nodal-ring configurations are obtained by WannierTools package~\cite{wu2018wanniertools}.

\section{Results and discussions}
\subsection{TB models}
A minimal five-band TB model is constructed for the stacked honeycomb-distorted kagome lattice, with $p_z$ orbital at Wyckoff position 1a (atomic coordinate of $(1/3, 2/3, 1/2)$, forming the honeycomb lattice), $d_z^2$ orbital at Wyckoff position 3f (atomic coordinate of $(x, 0, 0)$, forming the distorted kagome lattice), in the non-SOC case: 

\begin{equation}
\begin{aligned}
H_{nSOC} &= \sum_{i\in \alpha,\beta} \epsilon_i c_i^{\dag}c_i+ \sum_{<i,j> \in \alpha} t_1 c_i^{\dag}c_j + \sum_{<i,j>\in\beta} t_2 c_i^{\dag}c_j \\
&+ \sum_{i\in \alpha} t_{z1} c_i^{\dag}c_i + \sum_{i\in \beta} t_{z2} c_i^{\dag}c_i + \sum_{i\in\alpha,j\in\beta} t_z c_i^{\dag}c_j + h.c. \label{eq1} 
\end{aligned}
\end{equation}

where $\alpha$ and $\beta$ are the labels for kagome lattice and honeycomb lattice. $t_1$ and $t_2$ are the hopping between nearest sites in two sublattices, i.e., the intralayer hoppings in each sublattice, as illustrated in Fig.~\ref{fig3} (a). $t_{z}$ is the interlayer hopping between two sublattices. $t_{z1}$ and $t_{z2}$ are the hopping for the same sublattices in different unit cells along the stacking direction, respectively. The Hamiltonian can be expressed in the matrix form below: 
\begin{equation}
H_{nSOC} = \begin{pmatrix}
H_h & H_{hk} \\
H_{hk}^{\dag} & H_k \\
\end{pmatrix}
\end{equation}

\begin{widetext}
\begin{equation}
\centering
H_h = \begin{pmatrix}
    e_2 + 2t_{z2} \text{cos } k_z & t_2 e^{-1/3i(2k_x+k_y)} (1+e^{ik_x}+e^{i(k_x+k_y)}) \\
    t_2 e^{-1/3i(k_x+2k_y)} (1+e^{ik_y}+e^{i(k_x+k_y)}) & e_2 + 2t_{z2} \text{cos } k_z
\end{pmatrix}
\end{equation}

\begin{equation}
H_{hk} = \begin{pmatrix}
     t_z (-1+e^{i k_z}) e^{i(\mathbf{d_1} \cdot \mathbf{k})} & t_z (-1+e^{i k_z}) e^{i(\mathbf{d_2} \cdot \mathbf{k})}  & t_z (-1+e^{i k_z}) e^{i(\mathbf{d_3} \cdot \mathbf{k})}  \\
     t_z (-1+e^{i k_z}) e^{i(\mathbf{d_4} \cdot \mathbf{k})}  & t_z (-1+e^{i k_z}) e^{i(\mathbf{d_5} \cdot \mathbf{k})}  & t_z (-1+e^{i k_z}) e^{i(\mathbf{d_6} \cdot \mathbf{k})} 
\end{pmatrix}
\end{equation}

\begin{equation}
H_k = \begin{pmatrix}
    e_1 + 2t_{z1}\text{cos } k_z & t_1 e^{-i \mathbf{A}} (1+e^{i(k_x+k_y)}) & t_1 e^{-i \mathbf{B}} (1+e^{ik_y}) \\
    * & e_1 + 2t_{z1}\text{cos } k_z & t_1 e^{-i \mathbf{C}} (1+e^{ik_x}) \\
    * & * & e_1 + 2t_{z1}\text{cos } k_z
\end{pmatrix}
\end{equation}

\end{widetext}
where $\mathbf{d_1}$ - $\mathbf{d_6}$ are the vectors connecting the nearest neighbor kagome lattice and honeycomb lattice sites ($\mathbf{d_1} = (x-1/3, 1/3, -1/2)$, $\mathbf{d_2} = (-1/3, x-2/3, -1/2)$, $\mathbf{d_3} = (2/3-x, 1/3-x, -1/2)$, $\mathbf{d_4} = (x-2/3, -1/3, -1/2)$, $\mathbf{d_5} = (1/3, x-1/3, -1/2)$, $\mathbf{d_6} = (1/3-x, 2/3-x, -1/2)$) and $\mathbf{A}$ - $\mathbf{C}$ are the vectors connecting the nearest kagome lattice sites ($\mathbf{A} = (x, 1-x, 0)$, $\mathbf{B} = (2x-1, x, 0)$, $\mathbf{C} = (x, 2x-1, 0)$).

\begin{figure*}
  \centering
  \includegraphics[width=1.0\textwidth]{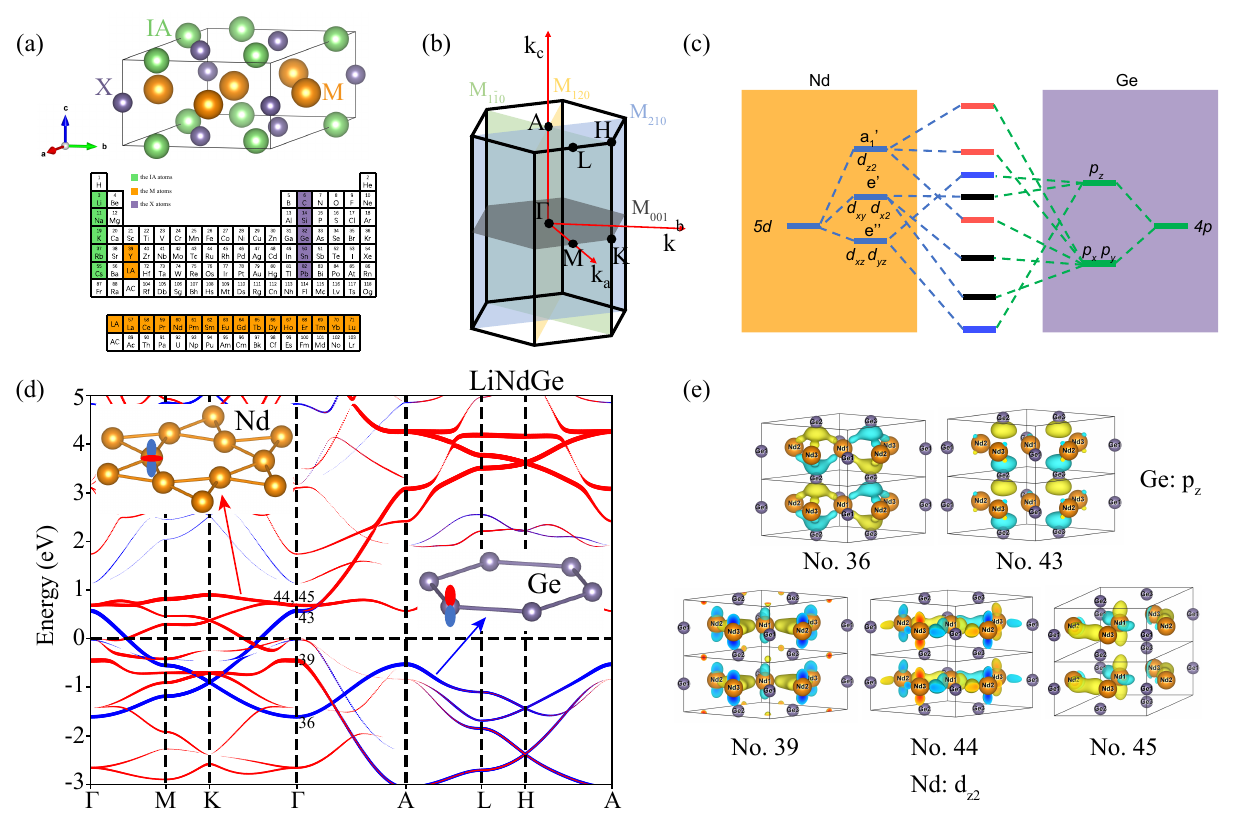}
  \caption{(a) Crystal structure of the IAMX family. The green, yellow, and purple balls represent IA, M, and X atoms, respectively. (b) BZ for the IAMX family, where the light green, light yellow, light blue, and dark gray planes are the $M_{1-10}$, $M_{120}$ and $M_{210}$ and $M_{001}$ planes, respectively. (c) The energy splitting in the bonding process for the orbitals of Nd and Ge atoms in the crystal around the Fermi level. (d) The band structure of LiNdGe in the non-SOC case, with the projections of Ge $p_z$ orbital and Nd $d_{z^2}$ orbital labeled in blue and red, respectively. (e) Wavefunctions for the energy bands at $\Gamma$ around the Fermi level, showing the compositions of Ge $p_z$ orbitals and Nd $d_{z^2}$ orbitals.\label{fig4}}
\end{figure*}

By fine-tuning the parameters of the TB model, nodal-ring semimetals with different configurations can be obtained, as the five different configurations shown in Figs.~\ref{fig3} (d-h). The node-ring topology is protected by mirror symmetry, thus ensuring their stable presence on the mirror planes~\cite{bzduvsek2016nodal,yang2018symmetry,fang2016topological,PhysRevX.8.031069,fang2015topological}. After considering SOC, these nodal-ring band crossings will open energy gaps, giving rise to the emergence of Weyl points at generic momenta, as shown by the red dots in Fig.~\ref{fig3}. 
We note that we only tune $t_1$ and $t_2$, i.e., the intralayer hopping strengths within two sublattices for simplicity, and keep the others fixed. Fig.~\ref{fig3} (b) shows the typical band structure without SOC depicted by solid blue
lines and with SOC depicted by dashed red lines. 
The nodal ring pinned on the $k_z=0$ plane shown Fig.~\ref{fig3} (c) is the band crossing with a filling number of 2, which will evolve into 6 pairs of Weyl points after SOC is taken into consideration. Figs.~\ref{fig3} (c-h) show the evolution of nodal-ring configuration as the parameter $t_2$ increases. The single nodal ring centered at $\Gamma$ will cross the BZ boundary and become several nodal rings centered at K as $t_2$ increases, and it gradually transforms into a nodal ring centered at the $\Gamma$ point in Fig.~\ref{fig3} (c), further evolving into multiple nodal-ring patterns depicted in Fig.~\ref{fig3} (e-h) as $t_2$ increases. Such an evolution process indicates the intralayer hopping strength in the sublattice will play an important role in the topological configuration.

\subsection{Scanning Materials with Stacked Kagome and Honeycomb Lattice}
By following the flowchart shown in Fig.~\ref{fig2} and the minimal five-band TB model, we discover that the IAMX family (IA = Alkali metal element, M = Rare earth metal element, X = Carbon group element) are ideal topological materials with stacked kagome and honeycomb lattice by scanning the ICSD. They belong to the space group $P$-62$m$ (No. 189), which has three equivalent mirror symmetries \emph{M}$_{1-10}$, \emph{M}$_{120}$ and \emph{M}$_{210}$, an inequivalent mirror symmetry \emph{M}$_{001}$ perpendicular to the $z$-axis, a threefold rotation symmetry $C_{3z}$ along $z$-axis and a twofold rotation symmetry $C_{2(110)}$. 
IA atoms occupy the Wyckoff position 3g, and M atoms are located at position 3f, forming a distorted kagome lattice, as depicted in Fig.~\ref{fig1} (c). 
X atoms occupy two nonequivalent Wyckoff positions 1a and 2d, and the latter ones form the honeycomb lattice, as shown in Fig.~\ref{fig1} (a). 
Thus, the crystal of the IAMX family can be regarded as composed of distorted kagome lattice and honeycomb lattice stacking along $z$ direction, as depicted in Fig.~\ref{fig4} (a). 
Fig.~\ref{fig4} (b) is the first BZ of the IAMX family, with $M_{1-10}$, $M_{120}$ and $M_{210}$ represented by light green, light yellow, and light blue planes, respectively, and the dark gray plane represents the non-equivalent (001) mirror plane. 
It is worth highlighting that the interlayer interactions of the IAMX family are chemical bonds rather than the van der Waals forces, which is distinct from previously discovered two-dimensional or quasi-two-dimensional materials featuring the kagome lattice. Thus, the interplay between the kagome lattice and honeycomb lattice may bring new physical properties for their 3D nature.

\subsection{Orbital Analysis for the IAMX family}
Since all the IAMX family materials share similar elemental components, band structures, and topology, we will take LiNdGe as an example for further analysis. In LiNdGe, $p$ orbital from Ge forms a honeycomb lattice, and $d$ orbital from Nd forms a distorted kagome lattice, and they will split into different bands under the crystal field, as shown in Fig.~\ref{fig4} (c). After the orbital hybridization between $d_{z^2}$ and $p_z$, bonding and antibonding bands can be obtained around the Fermi level, forming the band structure depicted in Fig.~\ref{fig4} (d).  
Fig.~\ref{fig4} (e) shows the orbital charge density distributions of different bands forming the kagome and honeycomb bands near the Fermi level, both the band structure and the orbital components can be captured by the aforementioned five-band TB model.

\subsection{Topological Analysis for the IAMX family}
To study the topological information of the IAMX family compounds, we calculate band structures, surface states, nodal-ring configuration, etc., through a high-throughput calculation, with SOC taken into account or not, respectively. 
298 ideal topological materials in the IAMX family are proceeded for further analysis. 
Within this dataset, we categorize the configurations of nodal rings into four main groups, as shown in Fig.~\ref{fig5}: (i) multiple nodal rings centered at K point on $k_z$ = 0 plane; (ii) a closed nodal ring located on the $k_z$ = 0 or $\pi$ plane centered at $\Gamma$ or A; (iii) a nodal-chain composed by six nodal rings on the mirror planes and an additional nodal ring on $k_z$ = 0 plane centered at $\Gamma$; (iv) three nested nodal rings on the mirror planes, accompanied by a nodal line at the intersection of the $\Gamma$-Z path. 
These four types of nodal-ring configuration highly depend on the chemical rule of the IAMX family, which will be discussed in detail below. 

\subsection{Chemical Rules for The Topology}
\subsubsection{Lattice Constant and Bonding strength}

\begin{figure}
  \centering
  \includegraphics[width=0.5\textwidth]{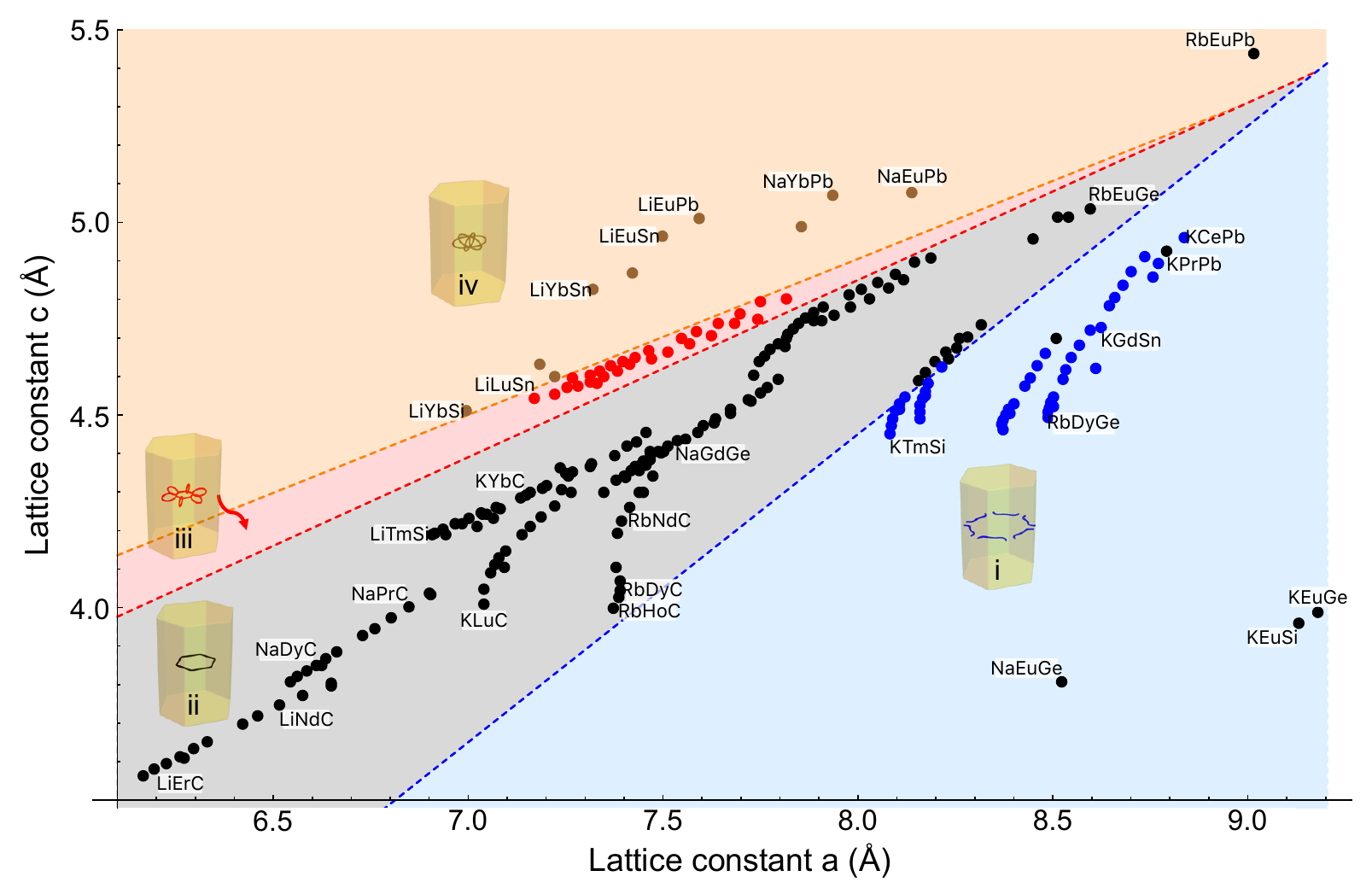}
  \caption{The distribution of four nodal-ring configurations under different lattice constants $a$ and $c$ for the IAMX family. Different nodal-ring configurations are labeled in different colors, separated by linear slashes. 
\label{fig5}}
\end{figure}

To study the influence of the overall bonding strength on the topological configuration, we categorize the distributions of 298 IAMX family materials based on their lattice constants $a$ and $c$ of the compounds in Fig.~\ref{fig5}, with type (i)-(iv) node-ring configurations fully separated in four regions, denoted by blue, black, red, and brown dots, respectively. We notice that different nodal-ring configurations can be separated by three linear dashed lines, showing a strong correlation between the topological configuration and the lattice constants.

Lattice constant $a$ corresponds to the parameters $t_1$ and $t_2$ in our TB model since it exerts influence on intralayer hopping in kagome or honeycomb lattice, while lattice constant $c$ corresponds to the parameter $t_z$ since it specifically impacts the interlayer coupling. 
If $c$ is fixed and $a$ decreases in IAMX family materials, the increase of intralayer coupling will lead to a change of the nodal-ring configuration from the right to the left of the region in Fig.~\ref{fig5}. 
As $a$ decreases, type-(i) configuration, i.e., nodal rings centered at K points on the $k_z$ = 0 plane, shrink gradually and transform into type-(ii) nodal ring, i.e., a closed nodal ring centered at $\Gamma$. When the closed nodal ring on the $k_z$ = 0 plane gradually shrinks and disappears, six nodal rings emerge on the vertical mirror planes, forming type-(iii) node-ring configuration, progressively approaching the $k_z$ axis, and eventually converging into type-(iv) node-ring configuration, i.e., three nested nodal-rings. 
As discussed in the TB model part, different node-line configurations can be obtained by tunning the parameter $t_1$, which is consistent with the node-ring configuration transition from (i) to (iv) by decreasing the lattice constant $a$, in which process the bond strength in each sublattice will be enhanced.

\begin{figure}
  \centering
  \includegraphics[width=0.5\textwidth]{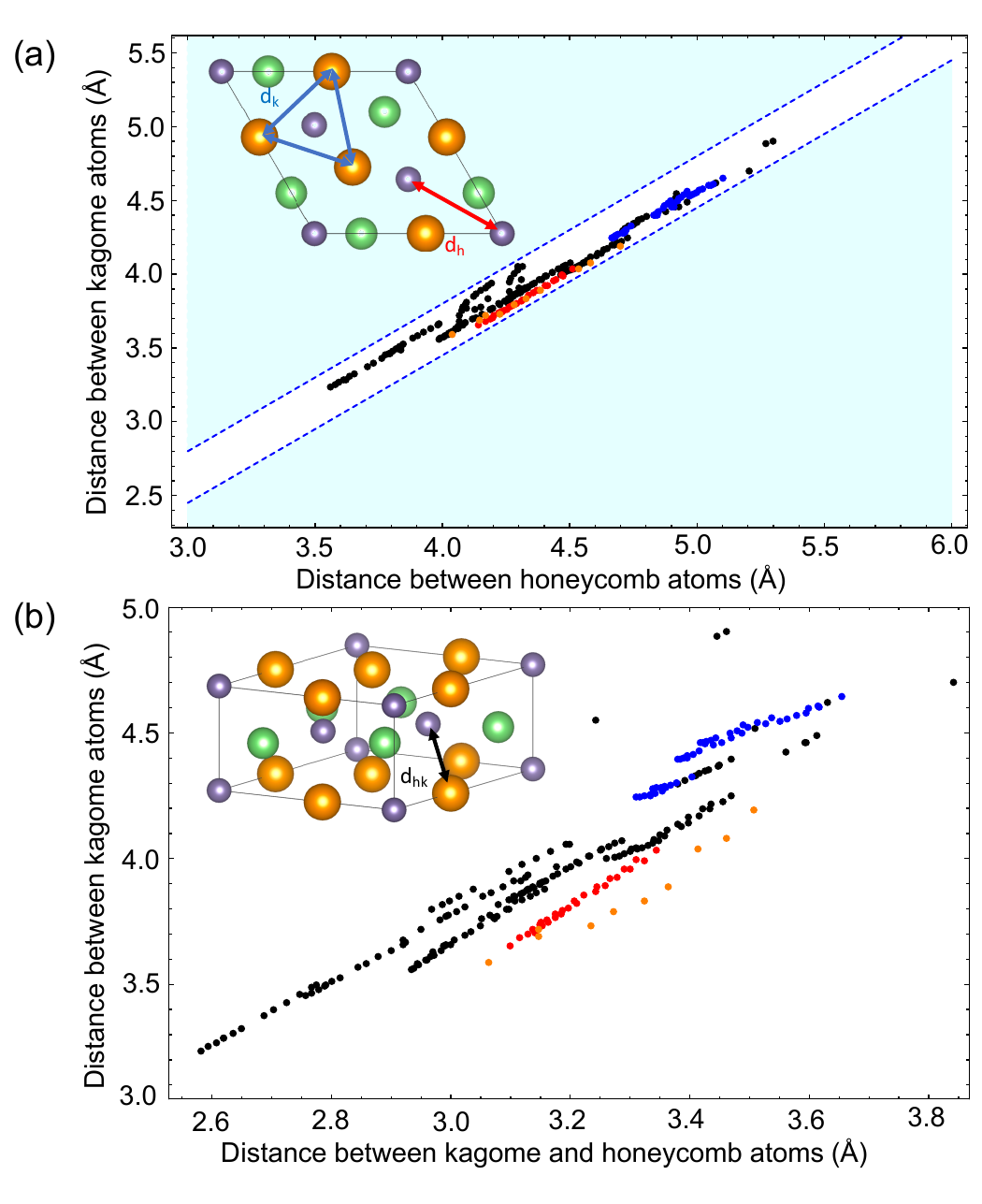}
  \caption{(a) The distribution of all IAMX family compounds with different intralayer atomic distances in both the honeycomb lattice ($d_h$) and kagome lattice ($d_k$). (b) The distribution of all IAMX family compounds with different interlayer atomic distances between the honeycomb and kagome lattice ($d_{hk}$). 
  Dots with different colors represent different node-ring configurations, as shown in Fig.~\ref{fig5}. 
  All the data is distributed in a linear way, showing a strong relation between the node-ring configuration and the atomic distances in the IAMX family. 
\label{fig6}}
\end{figure}

\subsubsection{Intralayer and Interlayer bond strength}
To explore the relationship between the topological configuration and the intralayer and interlayer chemical bonds, we study the node-ring configuration distribution of the IAMX family with different intralayer atomic distances in the honeycomb lattice ($d_h$), kagome lattice ($d_k$) and the interlayer lattice ($d_{hk}$), as shown in Fig.~\ref{fig6}. 

Figure~\ref{fig6} (a) illustrates the distribution relationship of all node-ring configurations based on the distance between X atoms forming the honeycomb lattice ($d_h$) and the distance between M atoms constituting the kagome lattice ($d_k$). All the data points are predominantly aligned along a straight line with the same gradient, which can be attributed to the linear dependence of both $d_h$ and $d_k$ on the lattice constant $a$ for different types of configurations. The intercepts of the lines, delineated by points representing distinct node-ring configurations in the diagram, exhibit correlation with the Wyckoff coordinates of the distorted kagome layer. Thus, different types of configurations can be distinguished through different magnitudes of intercepts, which highlights the significance of intralayer hopping as a key contributing factor influencing the formation of nodal-ring configurations.

Furthermore, we also analyze the distribution of topological configurations based on the distance between the atoms comprising the kagome and honeycomb lattice $d_{hk}$ and the distance between the atoms forming the kagome lattice $d_k$, as depicted in Fig.~\ref{fig6} (b). 
The interlayer hopping leads to more effective discrimination of various node-ring configurations, and simultaneously, a continuous transition links the type-(i) and type-(ii) node-ring configurations labeled by blue and black points, which is in alignment with Fig.~\ref{fig5}.

\subsubsection{The Role of Chemical Bond Strength for Nodal-line Configuration}

\begin{figure}
  \centering
  \includegraphics[width=0.5\textwidth]{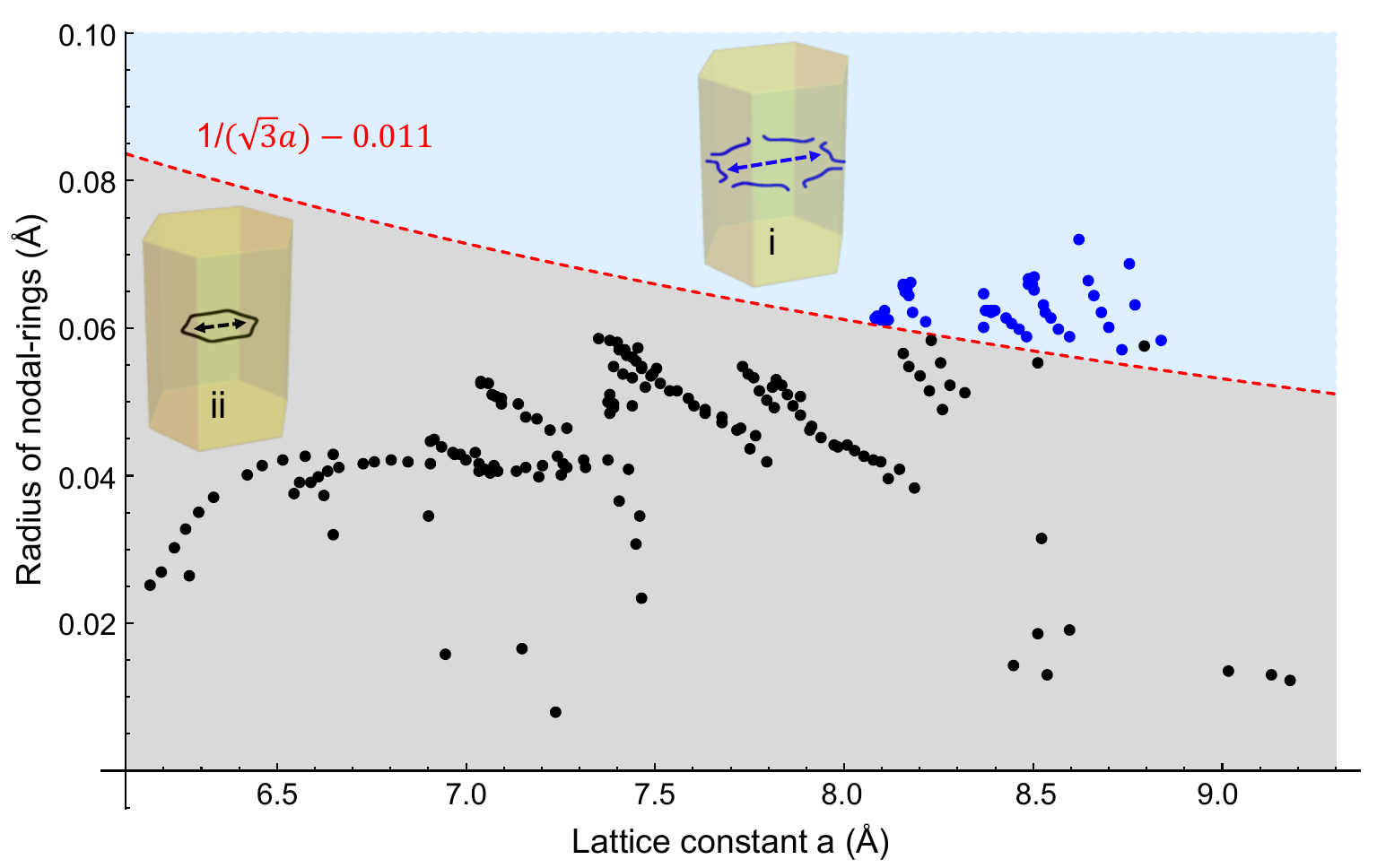}
  \caption{
  The distribution of IAMX family materials with the type-(i) configuration (blue points) and type-(ii) configuration (black points) with respect to the lattice constant $a$ and nodal-ring radius, which are distributed in the light blue and grey regions. 
  These two kinds of configurations are divided by a dashed line with a slope of $\frac{1}{\sqrt{3}a}$, where $a$ is the lattice constant, showing that these two configurations are similar in nature. 
\label{fig7}}
\end{figure}

\begin{figure}
  \centering
  \includegraphics[width=0.5\textwidth]{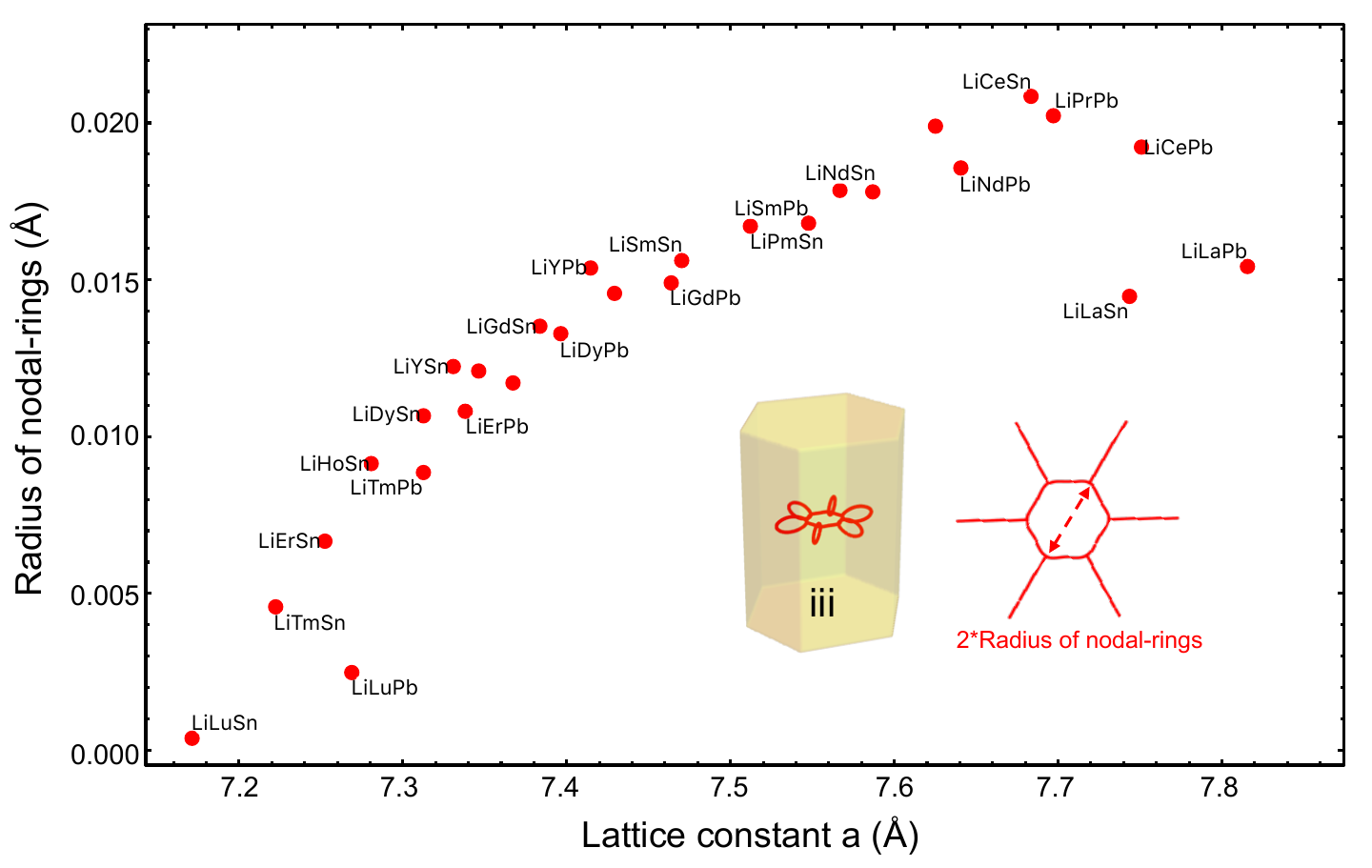}
  \caption{The distribution for the radius of type-(iii) configuration materials (red points in Fig.~\ref{fig5}) with respect to lattice constant $a$. 
\label{fig8}}
\end{figure}

Type-(i) configuration corresponds to a single closed nodal ring centered at $\Gamma$. If the radius becomes larger, the nodal ring will cross the BZ boundary and become six nodal rings centered at K, which will transform into the type-(ii) configuration. 
In order to have a further study on the effect role of chemical bond strength on node-ring configurations, we analyze the relationship between the configurations of type-(i) and type-(ii) nodal rings, i.e., the nodal-ring radius with respect to lattice constants $a$, as illustrated in Fig.~\ref{fig7}. 
These two configurations are well separated by a dashed line with a slope of 1/($\sqrt{3}$a), and they show a continuity at the boundary of these two regions. 
The increasing tendency for the radius of the nodal ring as the lattice constant $a$ increases on a whole shows that there is a strong correction between the topological configuration and the inplane lattice constant. Since the chemical bond strength in the kagome and honeycomb sublattices is also strongly corrected to the intralayer lattice constant, we conclude that the intralayer chemical bond strength plays an important role in modulating the topological configuration. 
However, we note that Fig.~\ref{fig7} shows a downward trend in each local part, which will be further explored in Part.~\ref{Electronegativity}.

Similarly, we also analyze the type-(iii) materials (red points in Fig.~\ref{fig5}) on the relationship between the node-ring radius and the lattice constant $a$, as depicted in Fig.~\ref{fig8}. Overall, their distribution exhibits an increasing tendency as $a$ increases, showing a strong correction between the node-ring radius and the intralayer bond strength, which matches the results shown in Fig.~\ref{fig7}. 

\begin{figure*}
  \centering
  \includegraphics[width=1.0\textwidth]{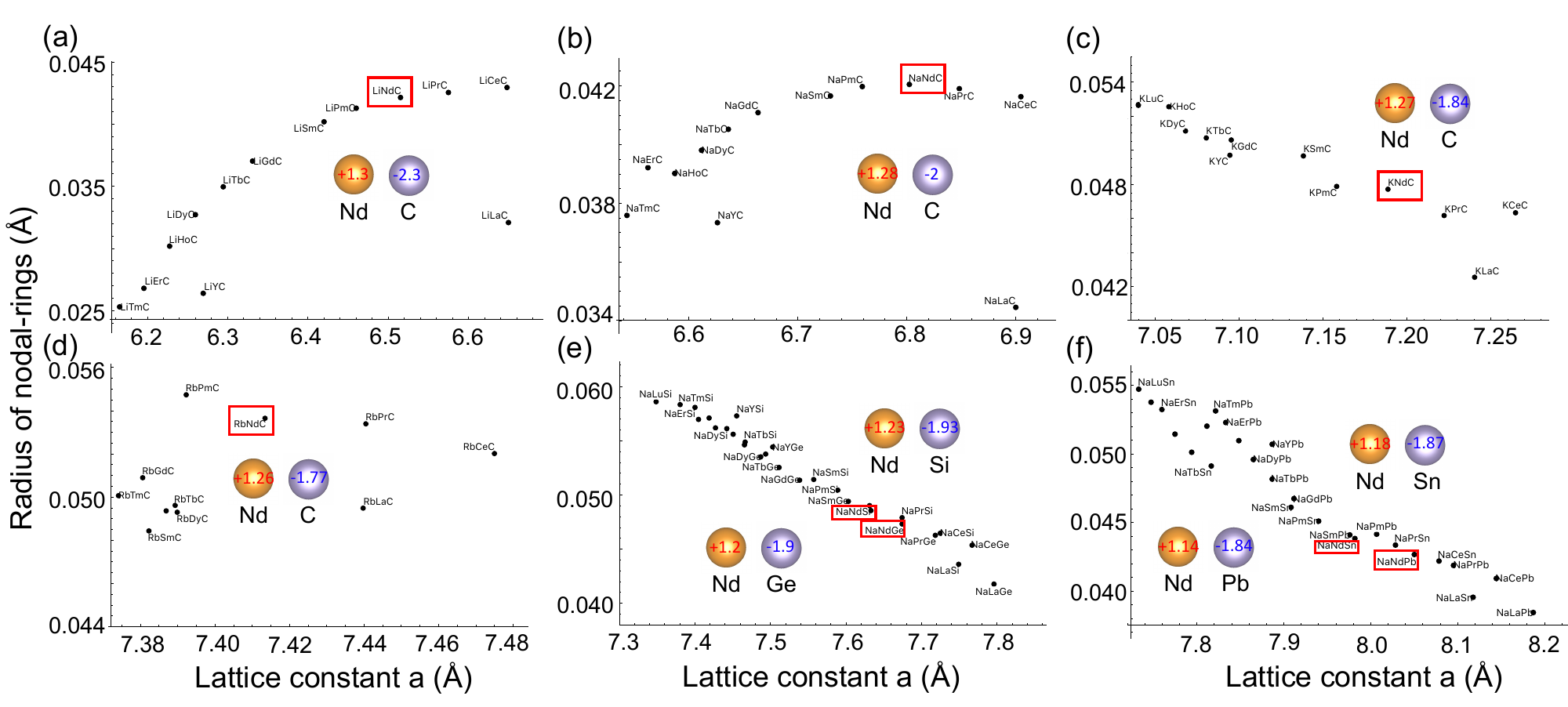}
  \caption{Further analysis of type-(ii) nodal-ring configuration in Fig.~\ref{fig5}, which corresponds to a single closed nodal-ring centered at $\Gamma$. The tendencies are different for the radius of the nodal ring and the lattice constant $a$ for different IA elements. 
\label{fig9}}
\end{figure*}

\subsubsection{The Influence of Electronegativity on Topology\label{Electronegativity}}

As we discussed above, even within the same node-ring configuration, the relationship between the node-ring radius and the lattice constant $a$ may locally show an opposite tendency compared to the global tendency, as shown in Fig.~\ref{fig7}. To have a deep insight into such phenomena, we take materials in type-(ii) configuration (black points) as an example, and further divide them into six groups, as shown in Fig.~\ref{fig9}, with X taking the carbon element and IA taking Li, Na, K, and Rb, respectively. 
Overall, the radius of the nodal ring and the lattice constant $a$ are linearly related, but with different slopes in these six groups. The slope of the linear dispersion decreases and even becomes zero when the IA element varies from Li to Rb. 
If we fix the IA element to Na, then the slope of the distribution gradually becomes negative, as shown in Fig.~\ref{fig9} (e-f). 

To gain further insight, we randomly selected one material from each group for further calculation. 
Our results reveal that each of the three elements in the IAMX family shows different valence states in each group. 
When the valence state of IV group element X is greater than -2, the slope of the distribution curve is positive. By contrast, when the valence of the X atom is less than -2, the slope is negative. While the valence state of X approaches -2, the slope of the curve tends to be around zero.
These results suggest that the electronegativity or electron affinity of the X element significantly influences the bonding strength, leading to variations in the node-ring radius and the topological configurations.

\subsection{Topological properties of LiNdGe with SOC}
Since materials within the IAMX family share similar crystal structures, band structures, and topological properties, we choose LiNdGe as an example to show the electronic structures and topological properties after taking SOC into consideration.

\begin{figure}
  \centering
  \includegraphics[width=0.5\textwidth]{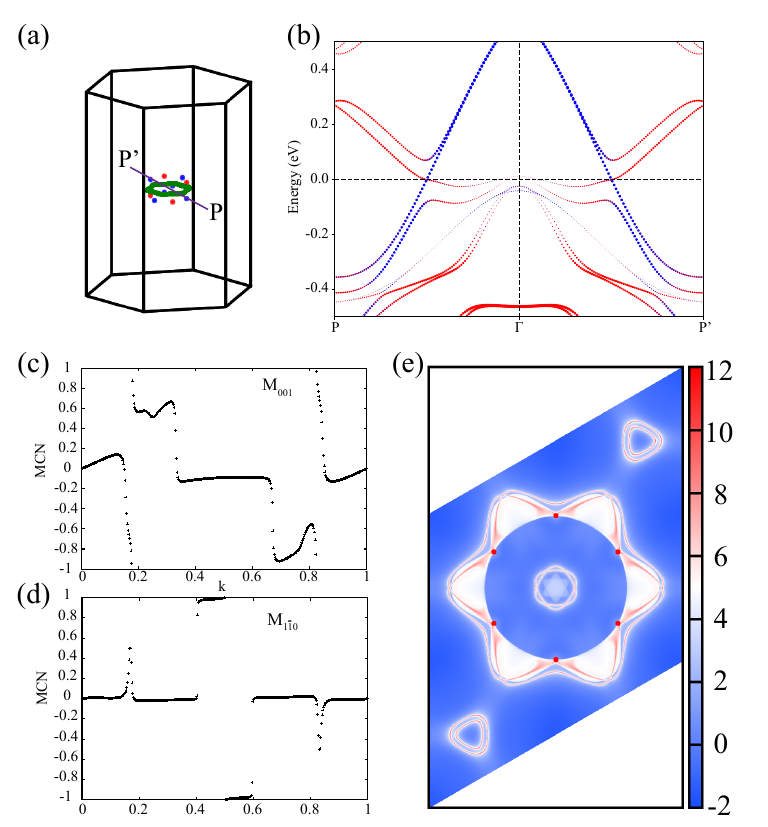}
  \caption{(a) Nodal-ring configuration (green line, without SOC) and the distribution of Weyl points (with SOC) in the BZ, where the blue and red dots denote the Weyl points with opposite chirality. (b) The band structure with SOC along the path $P-\Gamma-P'$. (c-d) The average position of the Wannier centers for occupied bands with mirror eigenvalue \emph{+i} for $M_{001}$ (c) and $M_{1-10}$ (d) planes, showing mirror Chern number of $-2$ and +1, respectively. (e) The Fermi surface including Weyl Fermi arcs along (001) direction for LiNdGe.\label{fig10}}
\end{figure}

The band structure of LiNdGe without SOC is depicted in Fig.~\ref{fig4} (d), where a nodal ring is located on the $k_z = 0$ plane, protected by the $M_{001}$ symmetry.  
When SOC is taken into account, all band crossings on the nodal ring are fully gapped, generating six pairs of Weyl points around the mirror plane, making LiNdGe a Weyl semimetal with energy bands contributed from the distorted kagome and honeycomb lattice. The distribution of these Weyl points is depicted in Fig.~\ref{fig10} (a), where blue/red dots represent Weyl points with negative/positive monopole charge, and they are constrained by the $M_{001}$ and $C_{3z}$ symmetries. 
Figure~\ref{fig10} (b) shows the band structure crossing two time-reversal-symmetry-related Weyl points along $P-\Gamma-P'$, and the energy of Weyl points is just 1.4 meV above the Fermi level, which means that IAMX family materials are ideal platforms to study physical properties induced by the Weyl fermions.

To have a better understanding of the topological information, we perform the mirror Chern numbers (MCN) calculations on the two non-equivalent $M_{001}$ and $M_{1-10}$ planes based on the Hamiltonian obtained from Wannier90. Fig.~\ref{fig10} (c-d) indicate the MCN of $M_{001}$ is 2, while that of $M_{1-10}$ is 1. 
According to the previous studies on TaAs~\cite{weng2015weyl,PhysRevX.5.031013}, if two non-parallel mirror planes have different MCNs, each of the four regions that these two mirror planes divide will contain an odd number of Weyl points. In this case, we find that there are three Weyl points in each of the four regions divided by $M_{001}$ and $M_{1-10}$, and 12 Weyl points in the whole BZ. Conversely, if the MCNs of the two non-parallel mirrors are the same, there will be 0 or an even number of Weyl points. Thus, in the regions divided by any two mirror planes among $M_{1-10}$, $M_{120}$, and $M_{210}$, there are 2 or 4 Weyl points in each region, thus there will also be 12 Weyl points in total. 

Fermi arc is one of the distinctive hallmarks of Weyl semimetals, and it connects the projected Weyl fermions with opposite chirality on the surface~\cite{weng2016topological,armitage2018weyl}. 
Thus, we perform a surface states calculation based on semi-infinite Green's function method for LiNbGe along (001) direction to demonstrate the existence of the Weyl fermions. 
12 Weyl points in LiNbGe project onto 6 momenta (marked by red dots) on the surface BZ and each of the projected points is composed of two Weyl points with opposite chirality. Figure~\ref{fig10} (e) shows that there are six Fermi arcs connecting two Weyl points that are the second nearest neighbor to each other.  

Here, we offer a perspective to understand the connection of Fermi arcs in terms of MCN, since the number of Fermi arcs is constrained by the mirror Chern numbers. 
The MCN of $M_{001}$ is 2, so there are two Fermi arcs cross through half of $M_{001}$ on the projection surface that is perpendicular to $M_{001}$. 
Likewise, the MCNs of $M_{1-10}$, $M_{120}$, and $M_{210}$ are all equal to 1, so there is one Fermi arc crossing through the half of these three mirror planes on the (001) surface, forming the Fermi arcs connection shown in Fig.~\ref{fig10} (e).

\section{Summary}
We study the topological properties and their chemical rules for systems with stacked kagome and honeycomb lattice, in both analytical ways based on the TB model and numerical ways based on the DFT calculation. 
The minimal five-band TB model shows that distinct node-ring semimetals and Weyl semimetals can be obtained by tuning the parameters, in the non-SOC case and SOC case, respectively. 
Combined with the TB model, we propose an algorithm for fast-scanning the materials with stacked kagome and honeycomb lattices in the ICSD and carry out a high-throughput calculation on their physical properties. 
298 ideal topological semimetals of the IAMX family are proposed, and their topological features are fully captured by the five-band TB model. 
Further analyses reveal that different topological configurations in IAMX family materials are determined by lattice constants, intralayer and interlayer bond strength, electronegativity, and so on. 
Our comprehensive study on the stacked kagome and honeycomb crystals paves the way for future studies on both topology-related physical properties in such systems and their interplay with CDW, superconductivity, axion insulators and other physical properties in kagome and honeycomb lattices.

\section{Acknowledgement} 
We acknowledge the support from the National Natural Science Foundation of China (Grant Nos. 12047503 and 12374165).

\appendix
\maketitle
\section{The DFT calculation results of some red dots}
\begin{figure}
  \centering
  \includegraphics[width=0.5\textwidth]{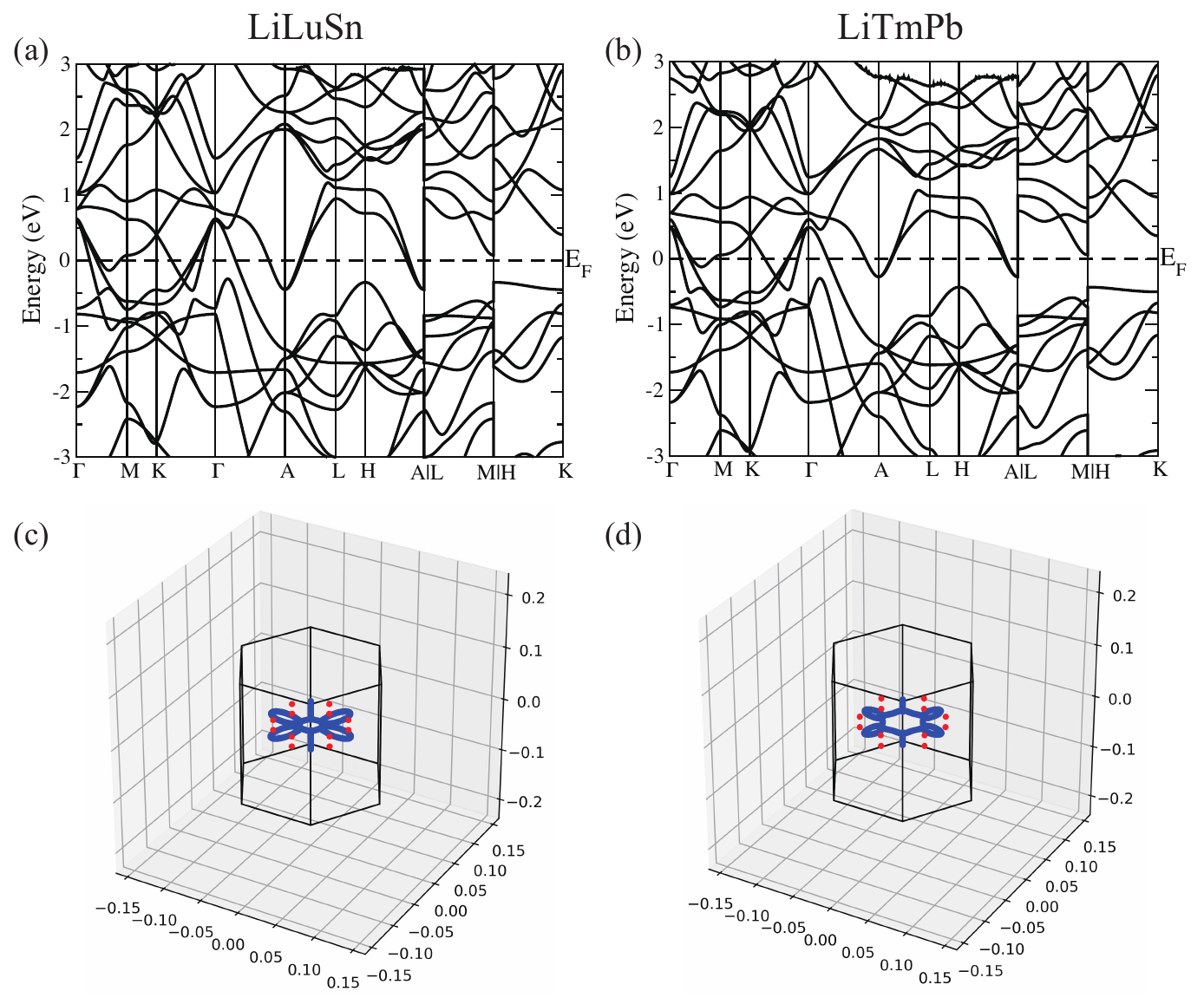}
  \caption{(a, b) The band structures of LiLuSn (a) and LiTmPb (b) without SOC. (c, d) The corresponding nodal-line configurations of LiLuSn (c) and LiTmPb (d).\label{SI1}}
\end{figure}

In this section, we show the DFT results of two materials from red dots (type-iii) as shown in Fig.~\ref{SI1}.

\section{The superconductivity of $\text{LiNdGe}$}
\begin{figure}
  \centering
  \includegraphics[width=0.5\textwidth]{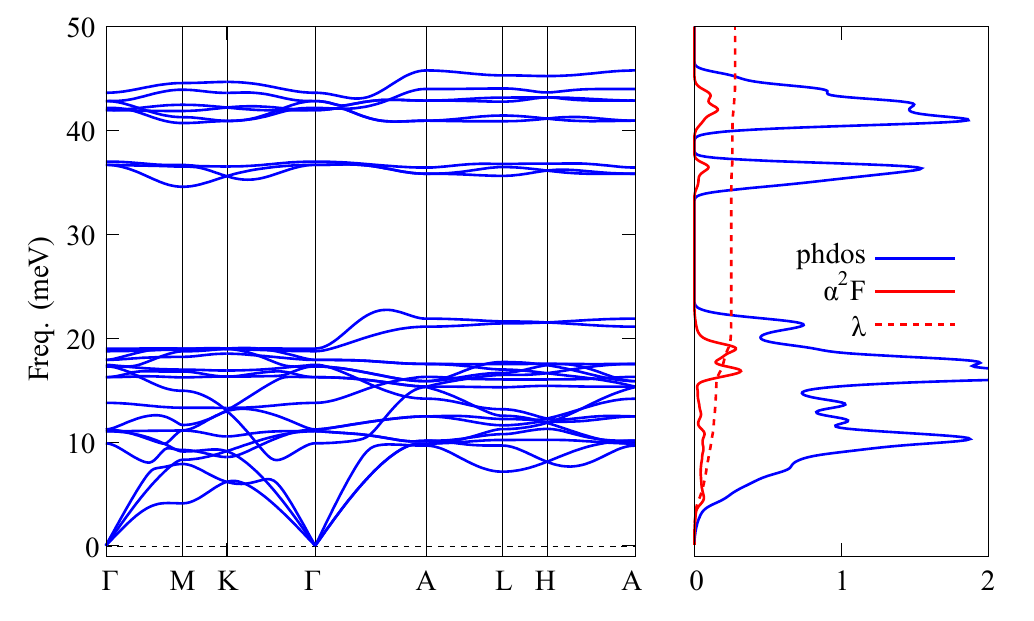}
  \caption{The phonon spectra, density of states, integrated electron-phonon coupling strength $\lambda$, and the $\alpha^2F$ spectrum of $\text{LiNdGe}$.\label{SI2}}
\end{figure}

We calculate the superconducting transition temperature ($\text{T}_{c}$) based on the BCS theorem ~\cite{BCS}. We perform the phonon and the electron-phonon coupling matrix using the \texttt{Quantum Espresso} package~\cite{Giannozzi_2009,Giannozzi_2017,giannozzi2020quantum}. The PBE functional ~\cite{perdew1996generalized} with PAW method ~\cite{blochl1994projector} is used with the $f$-shell electrons of Nd were treated as the semi-core electrons. The cutoff energy for wavefunctions is set as 60 Ry and a $7\times 7\times 11$ $k$-mesh is set to get converged total energy and charge density. We calculate the dynamic matrix at a $2\times 2\times 4$ $q$-point mesh. We use the \texttt{EPW} ~\cite{Giustino2007, Ponce2016} and the \texttt{Wannier90} package ~\cite{mostofi2014updated} to interpolate the electron-phonon coupling matrix to the full Brillouin zone (a $20\times 20\times 40$ fine mesh) and calculate the integrated electron-phonon coupling strength $\lambda$, the Eliashberg spectrum function $\alpha^2F$, as shown in Fig. ~\ref{SI2}, with phonon dispersion and DOS. According to the  McMillan-Allen-Dynes  formula~\cite{mcmillan1968transition,Allen_Dynes}, when we assume the effective Coulomb repulsion $\mu*=0.16$, the T$_c$ is less than $10^{-4}$ K, which rule out the possibility of the BCS superconductor due to the small electron-phonon coupling strength.

\bibliography{ref}

\end{document}